\documentclass[twocolumn,amssymb,showpacs,aps,prl]{revtex4-1}
\usepackage{}
\usepackage{stmaryrd}
\usepackage{amsmath}
\usepackage{amssymb}
\usepackage{float}
\usepackage{graphicx}
\usepackage{titletoc}
\usepackage{booktabs}
\usepackage{array, color}
\usepackage{tabularx}
\usepackage{latexsym,bm}
\usepackage{graphics}

\begin{document}


\title{Role of rotational coherence in  femtosecond-pulse-driven nitrogen ion lasing}

\author{Hongqiang Xie$^{1,2,\dag}$}
\author{Hongbin Lei$^{1,\dag}$}
\author{Guihua Li$^{1,3}$}
\author{Qian Zhang$^{1}$}
\author{Xiaowei Wang$^{1}$}
\author{Jing Zhao$^{1}$}
\author{Zhiming Chen$^{2}$}
\author{Jinping Yao$^{4,\ddag}$}
\author{Ya Cheng$^{5}$}
\author{Zengxiu Zhao$^{1}$}
\email{zhao.zengxiu@gmail.com}
\email{{\ddag} jinpingmrg@163.com}
\affiliation{$^1$Department of Physics, National University of Defense Technology, Changsha 410073, China\\
$^2$School of Science, East China University of Technology, Nanchang 330013, China\\
$^3$School of Science, East China Jiaotong University, Nanchang 330013, China\\
$^4$State Key Laboratory of High Field Laser Physics, Shanghai Institute of Optics and Fine Mechanics, Chinese Academy of Sciences, Shanghai 201800, China.\\
$^5$State Key Laboratory of Precision Spectroscopy, East China Normal University, Shanghai 200062, China\\
$\dag$These authors contributed equally to this work.}

\date{\today}

\begin{abstract}
We experimentally investigated the rotationally resolved polarization characteristics of N$_2^+$ lasing at 391 and 428 nm using a pump-seed scheme. By varying the relative angle between the linear polarizations of the pump and seed, it is found that the polarizations of the P and R branches of 391-nm lasing are counter-rotated. By contrast, both branches of 428-nm lasing remain polarized along the pump. The origin of the puzzled abnormal polarization characteristics is found based on a complete physical model that simultaneously includes the transient photoionization and the subsequent coupling among the electronic, vibrational and rotational quantum states of ions.
It suggests that the cascaded resonant Raman processes following ionization create negative coherence between the rotational states of $J$ and $J$+2 in the ionic ground state X$^2\Sigma_g^+(\nu=0)$, which leads to mirror-symmetrical polarization for the P and R branches of 391-nm lasing.  Both the experiment and theory indicate that the demonstrated rotational coherence plays an extremely pivotal role in clarifying the gain mechanism of N$_2^+$ lasing and opens up the route toward quantum optics under ultrafast strong fields.
\end{abstract}


\maketitle

Traditionally, quantum optics is based on weak light fields involving  only a few atomic states \cite{Meystre_QO}.
The emergence of high-intensity femtosecond lasers enables the excitation or ionization of molecules in a non-perturbative, non-equlibrium  fashion on a timescale of femtoseconds \cite{Mourou1,Strickland2}. Within the ultrashort time, the phases of the transition dipole moments of the violently excited molecules are hardly disturbed and quantum coherence among all the occupied states (including the continuum)  can be easily established \cite{Xia3,Steven4,Tamar5}. The strong field quantum optics, taking advantage of the readily available coherences created by femtosecond lasers, however has not yet received enough attention.


The fascinating nitrogen ion air lasing \cite{Yao6,Zhang7,Ando8,LiuYi2015, Zheng9,Britton10,Xie11}, created following strong field ionization, indicates that the quantum coherence remains even after the detachment of electrons. The lasing between the ionic states provides unparalleled advantages in terms of studying quantum effects and ultrafast dynamics of ions over other advanced spectroscopy techniques, such as high harmonic spectroscopy, holographic photoelectron spectroscopy and attosecond transient absorption spectroscopy \cite{Smirnova12,Villeneuve13,Goulielmakis14}.
However, a full understanding of the gain mechanism for N$_2^+$ air lasing has not been elucidated. It was initially suggested that the population inversion between the relevant energy levels of N$_2^+$ is responsible for the lasing based on the three-state coupling model \cite{Xu15,Yao16,Li17,Xie18,Zhang19}. Recent studies showed that the resonant coupling among the electronic-vibrational states of N$_2^+$ not only leads to population redistribution but also induces vibrational coherence and electronic coherence \cite{Chen20,Zhang21,Zhang22,Mysyrowicz23,Xu24}, which turn out to be vital for generation of N$_2^+$ lasing.
But the present investigation as shown below finds that the polarization characteristics depends on the rotational states which can not be explained by the previous conjectures. 


In this study, we found compelling evidence for the existence of rotational coherence in the vibrational ground state $\nu$=0 of the ionic ground state by measuring the abnormal polarization of the P and R branches of N$_2^+$ lasers. By considering the transient photoionization and cascaded resonant Raman processes, we demonstrate how the rotational coherence is built up and how it leads to lasing radiations and abnormal polarization properties, which helps clarify the controversial gain mechanism of N$_2^+$ lasers and the exploration of coherent brightened laser sources in air.  It deserves mentioning that the role of rotational coherence discussed herein is distinguished from macroscopic molecular alignment effect \cite{Zhang7,Zeng25,Azarm26,Larissian27,Xu31} which can merely modulate lasing gain at alignment delays. The non-zero value of rotational coherence in the vibrational state $\nu$=0 of ionic ground states arises from a cascaded resonant Raman process, which can also influence laser gain at nonalignment delays. In contrast, for a non-resonant case, the defined rotational coherence should be zero.The accessible rotational coherence with femtosecond lasers paves the way to study quantum optics effects based on ionic systems, including electromagnetically induced transparency (EIT), lasing without inversion (LWI), slow light effect, and optical storage.

The experimental setup is schematically illustrated in Fig. 1 similar to our previous work \cite{Li28}. Briefly, femtosecond laser pulses (1 kHz, 4 mJ, 800 nm, $\sim$35 fs) from a commercial Ti: sapphire laser system were split into two parts with a 7:3 beam splitter. The main pulse with an energy of 1.9 mJ was served as the pump. The other beam, after being frequency doubled by a 0.2-mm-thick $\beta$-barium borate crystal, was used as the seed. The measured seed pulse energy was $\sim$0.5 uJ. A Glan-Taylor prism (GT1) was placed in the seed-beam path to ensure that the seed pulse was linearly polarized. The polarization direction of the seed pulse could be varied by rotating a half-wave plate at 400 nm in the seed-beam path. The time delay between the pump and seed pulse was controlled by a motorized translation stage with a resolution of 670 attoseconds. The pump and seed pulses were combined with a dichromatic mirror (high reflectivity at 800 nm and high transmission at 400 nm) and then were focused with an $f=30$ cm lens into a gas chamber filled with nitrogen at the pressure of 12 mbar. Of note, both the polarizations and powers of the pump and seed after the dichromatic mirror were checked to preclude unnecessary effects brought by it. The generated forward N$_2^+$ lasing signals after collimated by an $f=25$ cm lens were collected with a system consisting of an integrating sphere and a spectrometer (Princeton Instrument HRS300). To examine the polarizations of the generated N$_2^+$ lasers, another Glan-Taylor prism (GT2) was placed before the integrating sphere when needed.

\begin{figure}
\includegraphics*[width=3.3in]{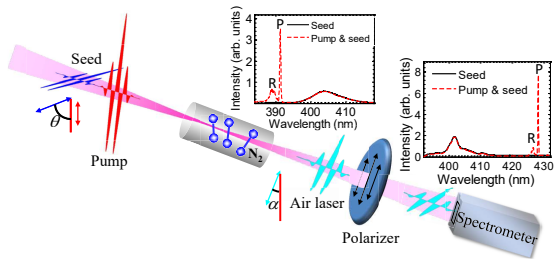}
\setlength{\abovecaptionskip}{-75pt}
\setlength{\belowcaptionskip}{-15pt}
\caption{(Color online)
Schematic diagram of the pump-seed setup. The insets are typical N$_2^+$ lasing spectra at 391 nm and 428 nm including rotationally-resolved P and R branches. $\theta$ represents the  angle between the linear polarizations of the pump and seed pulses, and $\alpha$ denotes the intersection angle between the polarization directions of the created N$_2^+$ lasers and the incident pump laser.
}\label{f1}
\end{figure}

Two typical lasing spectra at the pump-seed delay of $\sim$1 ps are plotted as insets in Fig. 1, showing that N$_2^+$ lasing at 391 and 428 nm can only be produced under the coexistence of the pump and seed. Both the lasing spectra are composed of P branch (rotational transition: $J \rightarrow J+1$) and R branch (rotational transition: $J \rightarrow J-1$). The involved lowest three electronic states X$^2\Sigma_g^+$, A$^2\Pi_u$ and B$^2\Sigma_u^+$ of N$_2^+$ are abbreviated as X, A, and B below.

\begin{figure}
\includegraphics*[width=3.3in]{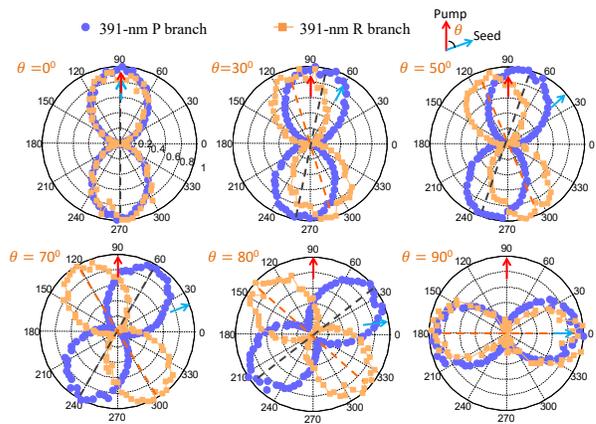}
\setlength{\abovecaptionskip}{-15pt}
\setlength{\belowcaptionskip}{-15pt}
\caption{(Color online)
The measured polarization states of the P and R branches of 391-nm lasing as a function of the polarization angle $\theta$. The blue and red arrows denote the polarization directions of the seed and the pump in each measurement, respectively. The dashed curves are for eye-guiding purposes.
}\label{f2}
\end{figure}

Figure 2 shows the polarization states of the P and R branches of 391-nm lasing (B($\nu$=0)$\rightarrow$ X($\nu$=0)) at different angle $\theta$ between the pump and the seed polarizations (indicated by red and blue respectively). Unexpectedly, it can be clearly seen that the P and R branches of 391-nm lasing are adversely deflected and their polarizations are nearly symmetric with respect to the pump polarization direction. This result indicates that for the electric-field component perpendicular to the pump polarization, the initial phases of the P branch and R branch lasing were reversed when the phases of the parallel component were kept constant. Moreover, the output N$_2^+$ laser polarization follows neither the polarization of the pump nor that of the seed except for the parallel and perpendicular cases (i.e., $\theta=0^\circ$ and $\theta=90^\circ$).
Note that the time delay between the pump and seed was fixed at a non-alignment delay of $\sim$3 ps to minimize the birefringent effect resulting from molecular alignment.

In addition, we performed the same polarization measurements for the N$_2^+$ lasing at 428 nm (B($\nu$=0)$\rightarrow$ X($\nu$=1)). The gas pressure was improved to 72 mbar to allow the 428-nm N$_2^+$ laser to be produced. The pump-seed delay was fixed at a non-alignment moment of $\sim$0.8 ps. We varied the polarization direction of the 800-nm pump using a half-wave plate at 800 nm placed in the pump-beam path in this case. The results are presented in Fig. 3. Surprisely, the polarizations of the P and R branches of 428-nm lasing are always the same for the different polarization angle $\theta$ and nearly comply with the polarization direction of the pump. Similar polarization behaviors have also been observed in the experiment of generating N$_2^+$ lasing with a 357-nm seed \cite{Miao29}, where the measured polarization of N$_2^+$ lasing followed the pump even though the pump polarization was adjusted to be vertical to that of the 357-nm seed.

\begin{figure}
\includegraphics*[width=3.3in]{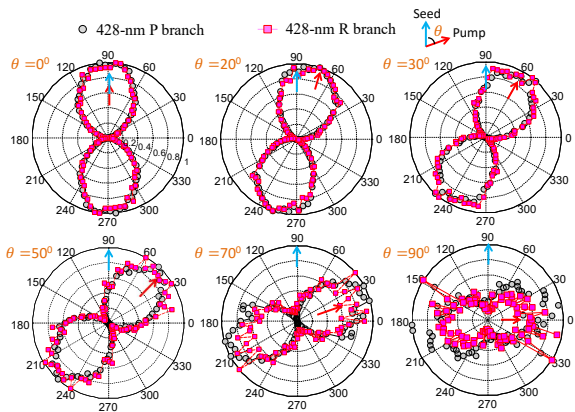}
\setlength{\abovecaptionskip}{-20pt}
\setlength{\belowcaptionskip}{-15pt}
\caption{(Color online)
Polarization measurements of the P branch and R branch lasing at 428 nm for different polarization angles $\theta$.
}\label{f3}
\end{figure}

Since the P and R branches of 391-nm lasing share the same upper rovibronic level, their different polarization characteristics could be caused by the difference in the lower rotational levels. To better understand the above experimental observations, we theoretically constructed a complete physical model that considers the contributions from electron, vibration and rotation on the basis of the ionization-coupling model proposed by us recently \cite{Zhang30}, by which we obtain the rotational coherence in the state X($\nu$=0, 1) after the interaction of an intense 800 nm pump laser.
Then, the interaction of the time-delayed seed pulse at $\sim$391 nm or $\sim$428 nm with the gain medium is solved using  the Maxwell's equation under the slow-varying envelope approximation (SVEA).

In our model, the three electronic states X, A and B are injected continuously through tunnel ionization during the pump.  For each electronic state, five vibrational states ($\nu$=0$\sim$4) and fifty rotational states ($J$=0$\sim$49) are included without loss of generality. Tunnel ionization rates taking into account the energy shift by the vibration are employed. The simultaneous coupling among these states following ionization is governed by the evolution of the ionic density matrix
\begin{equation}
\setlength{\abovedisplayskip}{3pt}
\setlength{\belowdisplayskip}{3pt}
\frac{d(\rho_{i\nu J}^+)}{dt}=-\frac{i}{\hbar}[{\rm H(t)},\rho_{i\nu J}^+(t)]+\left(\frac{d(\rho_{i\nu J}^+)}{dt}\right)_{\rm ionization}
\end{equation}
where i, $\nu$, and $J$ denote the electronic, vibrational and rotational state, respectively, and H is the ionic Hamiltonian. The last term $\left(\frac{d(\rho_{i\nu J}^+)}{dt}\right)_{\rm ionization}$ represents transient injection of ions due to ionization in the full laser pulse. We assume ionization induced coherence among ionic states to be zero, which means the non-diagonal elements are zero. The diagonal elements are written as:
\begin{equation}
\setlength{\abovedisplayskip}{3pt}
\setlength{\belowdisplayskip}{3pt}
\left(\frac{d(\rho_{i\nu J}^+)}{dt}\right)_{\rm ionization}=P_J\Gamma_{i\nu}\rho_0
\end{equation}
where $\rho_0$ is the time-decaying neutral population and $\Gamma_{i\nu}$ is the transient ionization rate for the respective electronic-vibrational states. $P_J$ is the Boltzmann distribution of the initiial rotational states for a given temperature T (T=298.5 K is used).

After the passage of the pump laser, the coherence $\sigma_{X(\nu=0, 1)}^{J, J+2}$ between the adjacent rotational states, $J$ and $J$+2 in the state X($\nu$=0, 1) can be retrieved from $\rho_{X(\nu=0, 1)}^{J, J+2}=\sigma_{X(\nu=0, 1)}^{J, J+2}e^{-i\omega t}$, where $\omega$ is the eigenfrequency difference of the rotational levels of $J$ and $J$+2.
At a delay of $\tau$ after the irradiation of the pump pulse, a weak seed pulse with a central wavelength at 391 nm or 428 nm is injected into the gain medium. Since the seed pulse is far away from resonance between the A-X coupling, it is reasonable that the seed pulse solely interacts with states B($\nu$=0) and X($\nu$=0, 1). Therefore, the Maxwell's equation using SVEA can be written as
\begin{equation}
\setlength{\abovedisplayskip}{3pt}
\setlength{\belowdisplayskip}{3pt}
(c\frac{\partial}{\partial z}+\frac{\partial}{\partial t})E_0(z, t)=i\Omega\rho_{BX}(z, t)
\end{equation}
where $E_0(z, t)$  is the Gaussian envelope of the propagating pulse with electric field $E(t)=E_0(t)e^{i(kz-\nu t)}$ and $\Omega=\left(\frac{2\nu N\mu_{BX}}{\varepsilon}\right)$ is the propagating constant. $N=10^{17} cm^{-3}$ is the population density of N$_2^+$; $\mu_{BX}$ is the transition dipole moment between the states B($\nu$=0) and X($\nu$=0, 1).
In the simulation, the pump polarization was set to $45^{\circ}$ against the molecular axis with a peak intensity of $\sim$2.5$\times 10^{14}$ W/cm$^2$ and duration of $\sim$35 fs. The seed pulse has a peak intensity of $\sim$2$\times 10^{10}$ W/cm$^2$ with duration of $\sim$80 fs. Fig.~4(a)  shows the final population distribution of the rotational sates of
 B($\nu$=0), X($\nu$=0) and X($\nu$=1) after the interaction with the pump laser. It can be seen that the population difference between the rotational level $J$ in the B($\nu$=0) and the rotational level $J$+1 or $J$-1 in the X($\nu$=0, 1) has been inverted. Interestingly, it can be seen from Fig. 4(b) that the rotational coherence $\sigma^{J, J+2}_X$ between the adjacent rotational states of the same parity, defined as the real part of $\sigma_{X(\nu=0, 1)}^{J, J+2}$, is negative for all the rotational states in the state X($\nu$=0), which signifies that the amplitude phases of the adjacent rotational levels of $J$ and $J$+2 have been reversed. In comparison, the coherence is nearly vanishing between the adjacent rotational levels $J$, $J$+2 in state X($\nu$=1). This finding will be shown crucial for the amplification of the seed pulse.

%

\begin{figure}
\includegraphics*[width=3.4in]{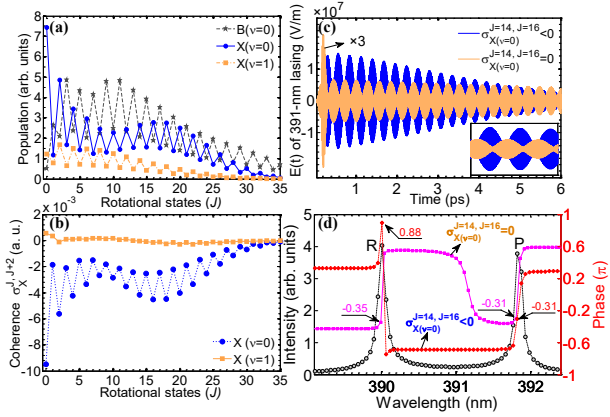}
\setlength{\abovecaptionskip}{-35pt}
\setlength{\belowcaptionskip}{-15pt}
\caption{(Color online)
(a) Simulated rotational distribution of the states B($\nu$=0), X($\nu$=0) and X($\nu$=1) after the interaction of the pump laser. (b) The coherence $\sigma^{J, J+2}_X$ among adjacent rotational states of $J$ and $J$+2 in states X($\nu$=0) and X($\nu$=1). (c) Calculated electric fields of 391-nm lasing from one single rotational level $J$=15 in B($\nu$=0) for the cases of $\sigma^{J=14, J=16}_{X(\nu=0)}<$0 and $\sigma^{J=14, J=16}_{X(\nu=0)}$=0. The inset is the close-up view for the time window of 0.4$\sim$1.2 ps. (d) Corresponding fast Fourier transformations for the electric fields in Fig. 4(c) to retrieve the spectral phase.
}\label{f4}
\end{figure}

In order to explain the polarization difference between 391-nm lasing and 428-nm lasing, we simulate the propagation of the injected seed pulse with time-delayed of 3ps in the gain medium of 1.5 mm length under the conditions of a rotational coherence $\sigma_{X(\nu=0)}^{J, J+2}<$0 and $\sigma_{X(\nu=0)}^{J, J+2}$=0 respectively.  For simplicity we consider  a $\Lambda$-type three-level system involving of the transitions from one single rotational state $J$ in the excited electronic state B to the $J$-1 and $J$+1 in the lower electronic state X. The isolated three-level system is justified when the mutual influence of the gain from different rotational levels is negligible during the propagation.
Taking the rotational lasing from B($\nu$=0, $J$=15) as an example. The rotational population and coherence of the involved levels can be extracted from Fig. 4(a) and 4(b) respectively. The blue curve in Fig. 4(c) shows the calculated electric field of 391-nm lasing from B($\nu$=0, $J$=15) to X($\nu$=0, $J$=14, 16). The amplified electric field can be written as $E(t)=\widetilde{E_P}(t)cos{(\omega_Pt+\phi_P)}$+$\widetilde{E_R}(t)cos{(\omega_Rt+\phi_R)}$ after removing the electric field of the seed pulse.
For comparison, we also calculated the laser field by assuming vanishing rotational coherence while keeping the same population inversion, as shown by the brown curve in Fig. 4(c).
The inset displays the electric fields in the time window of 0.4$\sim$1.2 ps. Apparently, the electric-field envelops for the two cases are almost opposite and the 391-nm lasing intensity is significantly improved under the circumstance of a negative coherence. Note that the simulated electric-field envelop for 428-nm lasing (not shown) is similar to that without rotational coherence.

To retrieve the spectral phase $\phi_P$, $\phi_R$ of both the P branch and R branch lasing, we performed fast Fourier transformations for the electric fields in Fig. 4(c). The results are plotted in Fig. 4(d). It can be seen that the phases of the spectral peaks for $\phi_P$ and $\phi_R$ are nearly the same for the case of $\sigma_{X(\nu=0)}^{J=14, J=16}$=0, which is -0.31$\pi$ and -0.35$\pi$, respectively. However, the phase difference of $\phi_P$ and $\phi_R$ reaches -1.19$\pi$ for a negative coherence, i.e., $\sigma_{X(\nu=0)}^{J=14, J=16}<$0.

\begin{figure}
\includegraphics*[width=3.4in]{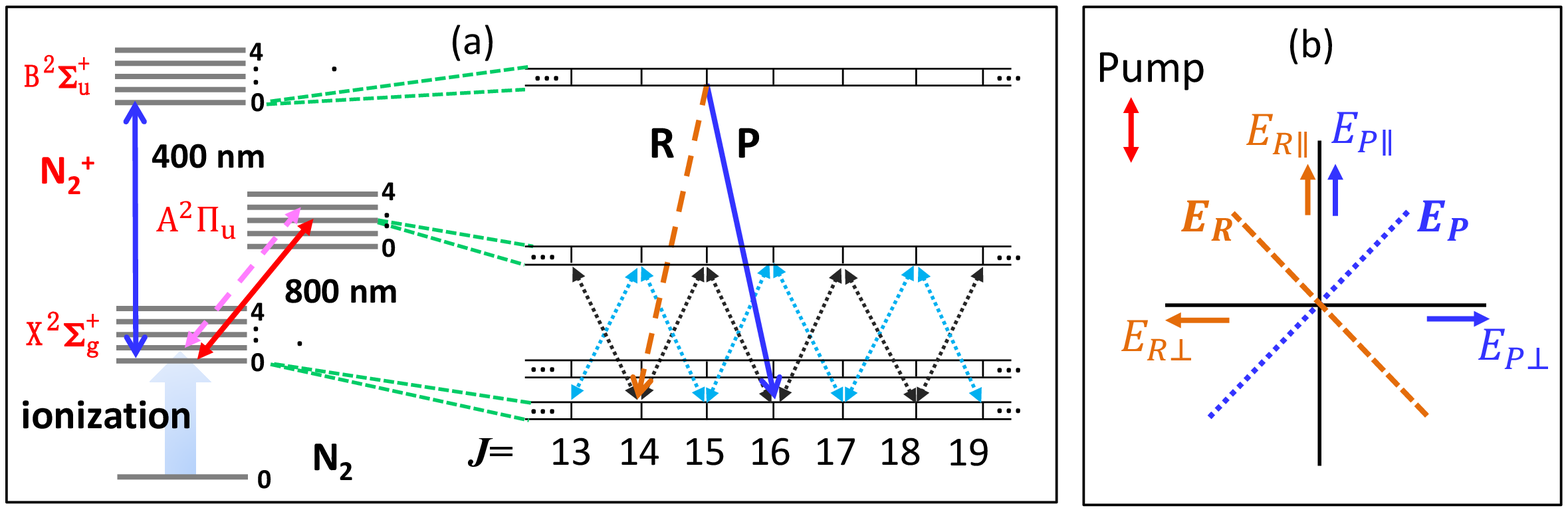}
\setlength{\abovecaptionskip}{-110pt}
\setlength{\belowcaptionskip}{-15pt}
\caption{(Color online)
(a) Preparation mechanism of the rotational coherence in the X($\nu=0$) state of N$_2^+$ via a cascaded Raman-like coupling between the X and A states. (b) Schematic polarizations of P and R branches of N${_2^+}$ 391-nm lasing.
}\label{f5}
\end{figure}

The above results can explain qualitatively the polarization features of 391-nm and 428-nm N$_2^+$ lasing. Since the ions created by transient ionization are mainly populated in the state X($\nu$=0), rather than in state X($\nu$=1) according to our previous calculations \cite{Zhang30}, the coupling between states X($\nu$=1) and A($\nu$=0$\sim4$) is too weak to induce rotational coherence in state X($\nu$=1). However, as shown in Fig. 5(a), resonant rotational coupling occurs during the electronic-vibrational coupling between states X($\nu=0$) and A($\nu=2$) and hence the rotational coherence in state X($\nu$=0) can be built up by the resonant cascaded Raman process. Note that the coupling of X-A belongs to vertical transitions and thus only along the direction perpendicular to the pump polarization can a macroscopic rotational coherence be prepared. Therefore, for N$_2^+$ lasing at 391 nm, its gain stems from two aspects, i.e., population inversion and rotational coherence. As shown in Fig. 5(b), the electric-field component perpendicular to the pump polarization is dominantly determined by the rotational coherence, which causes the the P and R branches of N$_2^+$ lasing can be nearly out of phase due to the negative coherence as depicted by the dotted lines with arrows in Fig. 5(a).
On the other hand, the parallel components are amplified based on the population inversion, leading to the same polarization for the P and R branches. Nevertheless, the gain origin of 428-nm lasing is the population inversion between states B($\nu=0$) and X ($\nu=1$), causing the P and R branches of 428-nm lasing to polarize in the same direction for all the angles $\theta$.

In conclusion, we have unambiguously revealed the rotational coherence in the vibrational state $\nu$=0 of the X state by experimentally measuring the polarizations of the ro-vibrationally resolved N$_2^+$ lasers.  We show that the strong field ionization along with the cascaded Raman coupling creates unique rotational coherence that leads to the difference polarization characteristics of the P and R branches lasing.  A $\Lambda$-type three-level model has been proposed to simulate the optical gain of the N$_2^+$ lasers. The consistency between the experimental observations and theoretical predictions confirms that the rotational coherence plays a pivotal role in N$_2^+$ laser gain, especially when the population is not fully inverted. Our findings shed new light on the physical mechanism of N$_2^+$ lasers and deepen the understanding of quantum coherence induced air lasing.  With a tunable femtosecond laser with a spectrum ranging from near-infrared to mid-infrared, strong field quantum optics fully taking advantage of the coherence of all degrees of freedom are within reach for other molecular systems.  Lastly, in terms of exploring applications with quantum coherence, the current unique ionic system can be advantageous over alkali atoms which are widely used to carry out the experiments of quantum optics previously. For example, in typical experiments of EIT, the demonstrated rotational coherence is expected to expand the bandwidth up to PHz scale, which is normally on a GHz scale in a usual coherent alkali atomic system.


\begin{acknowledgments}
This work is supported by the Major Research Plan of NSF (Grant No. 91850201), China National Key R\&D  Program (Grant No. 2019YFA0307703) and the National Natural Science Foundation of China (Grant Nos. 11822410, 61705034, 61605227 and 11704066).
\end{acknowledgments}


\begin{thebibliography}{99}

\bibitem{Meystre_QO}
Pierre Meystre and Murray Sargent III;  {\it Elements of Quantum Optics}, 3rd ed. (Springer-Verlag, Heidelberg, 1998).

\bibitem{Mourou1}
G. Mourou,
Rev. Mod. Phys. {\bf 91}, 030501 (2019).

\bibitem{Strickland2}
D. Strickland,
Rev. Mod. Phys. {\bf 91}, 030502 (2019).

\bibitem{Xia3}
H. Xia, A. A. Svidzinsky, L. Q. Yuan, C. Lu, S. Suckewer, and M. O. Scully,
Phys. Rev. Lett. {\bf 109}, 093604 (2012).

\bibitem{Steven4}
S. Yampolsky, D. A. Fishman, S. Dey, E. Hulkko, M. Banik, E. O. Potma and V. A. Apkarian,
Nat. Photonics {\bf 8}, 650 (2014).

\bibitem{Tamar5}
T. Seideman,
Phys. Rev. Lett. {\bf 83}, 4971 (1999).

\bibitem{Yao6}
J. Yao, B. Zeng, H. Xu, G. Li, W. Chu, J. Ni, H. Zhang, S. L. Chin, Y. Cheng and Z. Xu,
Phys. Rev. A {\bf 84}, 051802 (2011).

\bibitem{Zhang7}
H. Zhang, C. Jing, J. Yao, G. Li, B. Zeng, W. Chu, J. Ni, H. Xie, H. Xu, S. L. Chin, K. Yamanouchi, Y. Cheng and Z. Xu,
Phys. Rev. X {\bf 3}, 041009 (2013).

\bibitem{LiuYi2015}
Y. Liu, P.  Ding, G. Lambert, A. Houard, V. Tikhonchuk and A. Mysyrowicz,
Phys. Rev. Lett. {\bf 115}, 133203 (2015).

\bibitem{Ando8}
T. Ando, E. L\"{o}tstedt, A. Iwasaki, H.  Li, Y. Fu, S. Wang, H.  Xu and K. Yamanouchi,
Phys. Rev. Lett. {\bf 123}, 203201 (2019).

\bibitem{Zheng9}
W. Zheng, Z. M. Miao, L. Zhang, Y. Wang, C. Dai, A. Zhang, H. Jiang, Q. Gong and C. Wu,
J. Phys. Chem. Lett. {\bf 10}, 6598 (2019).

\bibitem{Britton10}
M. Britton, P. Laferriere, D. H. Ko, Z. Li, F. Kong, G. Brown, A. Naumov, C. Zhang, L. Arissian and P. B. Corkum,
Phys. Rev. Lett. {\bf 120}, 133208 (2018).

\bibitem{Xie11}
H. Xie, B. Zeng, G. Li, W. Chu, H. Zhang, C. Jing, J. Yao, J. Ni, Z. H. Wang, Z. Li and Y. Cheng,
Phys. Rev. A {\bf 90}, 042504 (2014).

\bibitem{Smirnova12}
O. Smirnova, Y. Mairesse, S. Patchkovskii, N. Dudovich, D. Villeneuve, P. Corkum and M. Ivanov,
Nature {\bf 460}, 972 (2009).

\bibitem{Villeneuve13}
D. M. Villeneuve, P. Hockett, M. J. J. Vrakking, H. Niikura, 
Science {\bf 356}, 110 (2017).

\bibitem{Goulielmakis14}
E. Goulielmakis, Z. H. Loh, A. Wirth, R. Santra, N. Rohringer, V. S. Yakovlev, S. Zherebtsov, T. Pfeifer, A. M. Azzeer, M. F. Kling, S. R. Leone and F. Krausz,
Nature {\bf 466}, 739 (2010).

\bibitem{Xu15}
H. Xu, E. L\"{o}tstedt, A. Iwasaki, and K. Yamanouchi,
Nat. Commun. {\bf 6}, 8347 (2015).

\bibitem{Yao16}
J. Yao, S. Jiang, W. Chu, B. Zeng, C. Wu, R.  Lu, Z.  Li, H.  Xie, G.  Li, C. Yu, Z.  Wang, H. Jiang, Q. Gong and Y. Cheng,
Phys. Rev. Lett. {\bf 116}, 143007 (2016).

\bibitem{Li17}
H.  Li, M.  Hou, H. Zang, Y. Fu, E. L\"{o}tstedt, T.
Ando, A. Iwasaki, K. Yamanouch and H. Xu,
Phys. Rev. Lett. {\bf 122}, 013202 (2019).

\bibitem{Xie18}
H. Xie, Q. Zhang, G. Li, X. Wang, L. Wang, Z. Chen, H. Lei and Z. Zhao,
Phys. Rev. A {\bf 100}, 053419 (2019).

\bibitem{Zhang19}
Y. Zhang, E. L\"{o}tstedt and K. Yamanouchi,
J. Phys. B: At. Mol. Opt. Phys. {\bf 52} 055401 (2019).

\bibitem{Chen20}
J. Chen, J. Yao, H. Zhang, Z. Liu, B. Xu, W. Chu, L. Qiao, Z. Wang, J. Fatome, O. Faucher, C. Wu and Y. Cheng,
Phys. Rev. A {\bf 100}, 031402 (2019).

\bibitem{Zhang21}
A. Zhang, M. Lei, J. Gao, C. Wu, Q. Gong and  H. Jiang,
Opt. Express {\bf 27}, 14922 (2019).

\bibitem{Zhang22}
A. Zhang, Q.  Liang, M.  Lei, L. Yuan, Y. Liu, Z.  Fan, X. Zhang, S.  Zhuang, C.  Wu, Q. Gong and H.  Jiang,
Opt. Express {\bf 27}, 12638 (2019).

\bibitem{Mysyrowicz23}
A. Mysyrowicz, R. Danylo, A. Houard, V. Tikhonchuk, X.
Zhang, Z. Fan, Q. Liang, S. Zhuang, L. Yuan and Y. Liu,
APL Photon. {\bf 4}, 110807 (2019).

\bibitem{Xu24}
B. Xu, J.  Yao, Y.  Wan, J.  Chen, Z.  Liu, F.  Zhang, W. Chu and Y. Cheng,
Opt. Express {\bf 27}, 018262 (2019).

\bibitem{Zeng25}
B. Zeng, W. Chu, G. Li, J. Yao, H. Zhang, J. Ni, C. Jing, H. Xie,
and Y. Cheng,
Phys. Rev. A {\bf 89}, 042508 (2014).

\bibitem{Azarm26}
A. Azarm, P. Corkum and P. Polynkin,
Phys. Rev. A {\bf 96}, 051401 (2017).


\bibitem{Larissian27}
L. Arissian, B. Kamer, A. Rastegari, D. M. Villeneuve, and J. C. Diels,
Phys. Rev. A {\bf 98}, 053438 (2018).
\bibitem{Xu31}
H. Xu, E. L\"{o}tstedt, T. Ando, A. Iwasaki and K. Yamanouchi,
Phys. Rev. A {\bf 96}, 041401 (2017).


\bibitem{Li28}
Z. Li, B. Zeng, W. Chu, H. Xie, J. Yao, G. Li, L. Qiao,
Z. Wang and Y. Cheng,
Sci. Rep. {\bf 6}, 21504 (2016).

\bibitem{Miao29}
Z. Miao, X. Zhong, L. Zhang, W. Zheng, Y. Gao, Y. Liu, H. Jiang, Q. Gong and C. Wu,
Phys. Rev. A {\bf 98}, 033402 (2018).

\bibitem{Zhang30}
Q. Zhang, H. Xie, G. Li, X. Wang, H. Lei, J. Zhao, Z. Chen, J.
Yao, Y. Cheng and Z. Zhao,
Commun. Phys. {\bf 3}, 50 (2020).
\end{thebibliography}
\end{document}